\documentstyle[12pt]{article}

\begin{document}
\title{GRAVITATIONAL RADIATION, VORTICITY AND THE ELECTRIC AND MAGNETIC
PART OF WEYL TENSOR}
\author{ L. Herrera$^{1}$\thanks{e-mail: laherrera@telcel.net.ve},   N. O.
Santos$^{2,3}$\thanks
{e-mail: nos@cbpf.br and santos@ccr.jussieu.fr} and J.
Carot$^{4}$\thanks{e-mail: jcarot@uib.es}\\
\small{$^1$Escuela de F\'{\i}sica, Facultad de Ciencias,} \\
\small{Universidad Central de Venezuela, Caracas, Venezuela.}\\
\small{$^2$Laborat\'orio Nacional de Computa\c{c}\~{a}o Cient\'{\i}fica,}\\
\small{25651-070 Petr\'opolis RJ, Brazil.}\\
\small{$^2$LERMA/CNRS-FRE 2460, Universit\'e Pierre et Marie Curie, ERGA,}\\
\small{Bo\^{\i}te 142, 4 Place Jussieu, 75005 Paris Cedex 05, France.}\\
\small{$^4$Departament de  F\'{\i}sica,}\\
\small{Universitat Illes Balears, E-07122 Palma de Mallorca, Spain}}
\maketitle
\begin{abstract}
The electric and the magnetic part of the Weyl tensor, as well as the
invariants obtained from them, are calculated for the Bondi vacuum metric.
One of the invariants vanishes identically  and the other  only exhibits contributions from terms of the Weyl tensor containing the static part of the field. It is shown that the
necessary and sufficient condition for the spacetime to be  purely electric is that such
spacetime be static. It is also shown that the vanishing of the electric
part implies Minkowski spacetime.
Unlike the electric part, the magnetic part does not contain contributions
from the static field. Finally a speculation about the link between the
vorticity of world lines of
observers at rest in a Bondi frame, and gravitational radiation, is 
presented.
\end{abstract}
\newpage
\date{}

\section{Introduction}
The study of the electric $E_{\alpha \beta}$ and magnetic $H_{\alpha
\beta}$ parts of Weyl tensor has atracted the attention of
researchers for many years (see \cite{Bel}--\cite{Ferrando} and
references therein).

Particularly intriguing is the eventual relationship of the magnetic part
of the Weyl tensor, with rotation \cite{glass, Bonnor}  and with
gravitational radiation \cite{Bel, Bruni, Bonnora, Dunsby, Maartens, Hogan}.

On the other hand, the link  has been established between  gravitational
radiation and vorticity of world--lines of observers at rest in a Bondi
space--time \cite{HHP, Valiente}.
Specifically, it has been shown that the leading term in the vorticity (in
an expansion of powers of $1/r$) is expressed through the news function in
such a way that it will vanish if
and only if there  is no news (no radiation). This suggests the possibility
of detecting gravitational waves by means of gyroscopes \cite{HHP, Felice}.

In order to delve deeper into these issues,  we shall calculate in this
work the electric and the magnetic part of the Weyl tensor, as well as the
two invariants obtained from
them, in the field of gravitational radiation.

 From the obtained expressions, it  follows that the vanishing of the
magnetic part, implies the vanishing of the news function and also
the vanishing of non--radiative but time dependent field, except for a very
peculiar class of solutions, called by Bondi ``non--natural, non--radiative
moving systems''. This result,
together with the known fact
\cite{Wade}  that  static Weyl metrics are purely electric, implies, if we
exclude by physical reasons the class of solutions mentioned before, that
the necessary and sufficient
condition for a Bondi spacetime to be purely electric is that such
space--time be static.

The vanishing of the electric part of the Weyl tensor is shown to imply
that the spacetime is Minkowski. Thus there is no purely magnetic vacuum
Bondi space--times, in agreement
with the conjecture that purely magnetic vacuum  space--times do no exist
\cite{Wade, Hadow, Bonnor, Bergh, Ferrando}.

It is also obtained that one of the invariants ($Q\equiv E^{\alpha
\beta}H_{\alpha \beta}$) vanishes identically whereas the other ($L\equiv
E^{\alpha \beta}E_{\alpha \beta}-H^{\alpha
\beta}H_{\alpha
\beta}$) has a leading term with contributions only from the coefficients in the expansion of the Weyl tensor  which contain the static part of the field. Coefficients containing purely
radiative and/or non--radiative but time dependent part of the field, do not enter in $L$.

Finally, we shall speculate  that the fact that gravitational radiation
produces vorticity of a time--like congruence, might be explained by a
mechanism similar to the one suggested to
explain the vorticity  of a time--like congruence in the field of a charged
static magnetic dipole \cite{bonleter}.

We shall carry out our calculations using the Bondi's formalism \cite{Boal}
which has, among other things, the virtue of providing a clear and precise
criterion for the existence of gravitational radiation (see also 
\cite{JaNe}).
Namely, if the news function is zero over a time interval,
then there is no radiation during that interval.

The formalism has as its main drawback \cite{BoNP}  the fact that
it is based on a series expansion which could not give closed
solutions and which raises unanswered questions about convergence
and appropriateness of the expansion.

However  we shall restrain ourselves to a region  sufficiently  far
from the source, so that  we shall need in our calculations only the
leading terms in the expansion of metric functions. Furthermore,
since the source is assumed to radiate during a finite interval,
then no problem of convergence appears \cite{Bo}.

A brief resume of Bondi's formalism is  given in the next section, together
with the expression of the vorticity of a time--like congruence of
observers at rest in a Bondi frame. In
section 3 we present the result of the calculations of the electric and  a
magnetic part as well as the invariants $Q$ and $L$, and in the last
section results are  discussed.

\section{The Bondi's formalism}
The general form of an axially and reflection symmetric asymptotically flat
metric given by Bondi is \cite{Boal}
\begin{eqnarray}
ds^2 & = & \left(\frac{V}{r} e^{2\beta} - U^2 r^2 e^{2\gamma}\right) du^2
+ 2 e^{2\beta} du dr \nonumber \\
& + & 2 U r^2 e^{2\gamma} du d\theta
- r^2 \left(e^{2 \gamma} d\theta^2 + e^{-2\gamma} \sin^2{\theta} 
d\phi^2\right)
\label{Bm}
\end{eqnarray}
where $V, \beta, U$ and $\gamma$ are functions of
$u, r$ and $\theta$.

We number the coordinates $x^{0,1,2,3} = u, r, \theta, \phi$ respectively.
$u$ is a timelike coordinate such that $u=constant$ defines a null surface.
In flat spacetime this surface coincides with the null light cone
open to the future. $r$ is a null coordinate ($g_{rr}=0$) and $\theta$ and
$\phi$ are two angle coordinates (see \cite{Boal} for details).

Regularity conditions in the neighborhood of the polar axis
($\sin{\theta}=0$), imply that
as $\sin{\theta}->0$
\begin{equation}
V, \beta, U/\sin{\theta}, \gamma/\sin^2{\theta}
\label{regularity}
\end{equation}
each equals a function of $\cos{\theta}$ regular on the polar axis.

The four metric functions are assumed to be expanded in series of $1/r$,
then using the field equations Bondi gets

\begin{equation}
\gamma = c r^{-1} + \left(C - \frac{1}{6} c^3\right) r^{-3}
+ ...
\label{ga}
\end{equation}
\begin{equation}
U = - \left(c_\theta + 2 c \cot{\theta}\right) r^{-2} + \left[2
N+3cc_{\theta}+4c^2 \cot{\theta}\right]r^{-3}...
\label{U}
\end{equation}
\begin{eqnarray}
V & = & r - 2 M\nonumber \\
& - & \left( N_\theta + N \cot{\theta} -
c_{\theta}^{2} - 4 c c_{\theta} \cot{\theta} -
\frac{1}{2} c^2 (1 + 8 \cot^2{\theta})\right) r^{-1} + ...
\label{V}
\end{eqnarray}
\begin{equation}
\beta = - \frac{1}{4} c^2 r^{-2} + ...
\label{be}
\end{equation}
where $c$, $C$, $N$ and $M$ are functions of $u$ and $\theta$, letters as
subscripts denote derivatives, and

\begin{equation}
4C_u = 2 c^2 c_u + 2 c M + N \cot{\theta} - N_\theta
\label{C}
\end{equation}

The three functions $c, M$ and $N$ are further
related by the supplementary conditions
\begin{equation}
M_u = - c_u^2 + \frac{1}{2}
\left(c_{\theta\theta} + 3 c_{\theta} \cot{\theta} - 2 c\right)_u
\label{Mass}
\end{equation}
\begin{equation}
- 3 N_u = M_\theta + 3 c c_{u\theta} + 4 c c_u \cot{\theta} + c_u c_\theta
\label{N}
\end{equation}

In the static case $M$ equals the mass of the system whereas $N$ and $C$
are closely related to the dipole and quadrupole moment respectively.

Next, Bondi defines the mass $m(u)$ of the system as
\begin{equation}
m(u) = \frac{1}{2} \int_0^\pi{M \sin{\theta} d\theta}
\label{m}
\end{equation}
which by virtue of (\ref{Mass}) and (\ref{regularity}) yields
\begin{equation}
m_u = - \frac{1}{2} \int_0^\pi{c_u^2 \sin{\theta} d\theta}
\label{muI}
\end{equation}

Let us now recall the main conclusions emerging from  Bondi's approach.
\begin{enumerate}
\item If $\gamma, M$ and $N$ are known for some $u=a$(constant) and
$c_u$ (the news function) is known for all $u$ in the interval
$a \leq u \leq b$,
then the system is fully determined in that interval. In other words,
whatever happens at the source, leading to changes in the field,
it can only do so by affecting $c_u$ and viceversa. In the
light of this comment the relationship between news function
and the occurrence of radiation becomes clear.
\item As it follows from (\ref{muI}), the mass of a system is constant
if and only if there are no news.
\end{enumerate}

Now, for an observer at rest in the frame of (\ref{Bm}), the four-velocity
vector has components
\begin{equation}
u^{\alpha} = \left(\frac{1}{A}, 0, 0, 0\right)
\label{fvct}
\end{equation}
with
\begin{equation}
A \equiv \left(\frac{V}{r} e^{2\beta} - U^2 r^2 e^{2\gamma}\right)^{1/2}.
\label{A}
\end{equation}

Then, it can be shown that for such an observer the vorticity vector may be
written as (see \cite{HHP} for details)
\begin{equation}
\omega^\alpha = \left(0, 0, 0, \omega^{\phi}\right)
\label{oma}
\end{equation}
with
\begin{eqnarray}
\omega^{\phi} & = & -\frac{e^{-2\beta}}{2 r^2 \sin{\theta}} \left[ 2
\beta_\theta e^{2\beta} - \frac{2 e^{2\beta} A_\theta}{A}
- \left(U r^2 e^{2\gamma}\right)_r  \right. \nonumber \\
  & + & \left. \frac{2 U r^2 e^{2\gamma}}{A} A_r +
\frac{e^{2\beta}\left(U r^2 e^{2\gamma}\right)_u}{A^2} - \frac{Ur^2
e^{2\gamma}}{A^2} 2 \beta_u e^{2\beta} \frac{}{} \right] \label{om3}
\end{eqnarray}
and for the absolute value of $\omega^\alpha$ we get
\begin{eqnarray}
\Omega & \equiv & \left(- \omega_\alpha \omega^\alpha\right)^{1/2} =
  \frac{e^{-2\beta -\gamma}}{2 r}
\left[2 \beta_\theta e^{2\beta} - 2 e^{2\beta} \frac{A_\theta}{A}
  -  \left(U r^2 e^{2\gamma}\right)_r\right.
\nonumber \\
  & + & \left. 2 U r^2 e^{2\gamma} \frac{A_r}{A}
  +  \frac{e^{2\beta}}{A^2} \left(U r^2 e^{2\gamma}\right)_u
  - 2 \beta_u \frac{e^{2\beta}}{A^2} U r^2 e^{2\gamma}
\right] \label{OM}
\end{eqnarray}
Feeding back (\ref{ga}--\ref{be}) into (\ref{OM}) and
keeping only the two leading terms, we obtain
\begin{eqnarray}
\Omega & = &-\frac{1}{2r} ( c_{u \theta}+2 c_u \cot \theta) \nonumber \\
&  & +\frac 1{r^2} \left[ M_{\theta}-M (c_{u \theta}+2 c_u \cot 
\theta)-c c_{u
\theta}+6 c c_u \cot \theta+2 c_u c_{\theta} \right]
\label{Om2}
\end{eqnarray}

Therefore, up to order $1/r$, a gyroscope at rest in (\ref{Bm}) will
precess as long as the system radiates ($c_{u} \not= 0$). Observe that if
\begin{equation}
c_{u\theta} + 2 c_u \cot{\theta} = 0
\label{if}
\end{equation}
then
\begin{equation}
c_u = \frac{F(u)}{\sin^2{\theta}}
\label{cu}
\end{equation}
which implies
\begin{equation}
F(u) = 0  \Longrightarrow c_u = 0
\label{F}
\end{equation}
in order to insure regularity conditions, mentioned above, in the
neighbourhood of the polar axis ($\sin{\theta} = 0$) . Thus the
leading term in (\ref{Om2}) will vanish if and only if $c_u = 0$.

The order $1/r^2$  contains, beside the
terms involving $c_u$, a term not involving news, namely
$M_{\theta}$. This last term represents the class of
non-radiative motions discussed by Bondi \cite{Boal} and may be thought 
of as
corresponding to the tail of the wave, appearing after the radiation process
\cite{BoNP}.

Let us now assume that initially (before some $u = u_0 =
$constant) the system is static, in which case
\begin{equation}
N_{u}=c_u = 0
\label{static}
\end{equation}
which implies, because of (\ref{N}),
\begin{equation}
M_{\theta} = 0
\label{staticII}
\end{equation}
\noindent
and $\Omega = 0$ (actually, in this case $\Omega=0$ at any order) as 
expected
for  a static field ( for the electrovacuum case however, this may change
\cite{bonleter}). Then let us suppose that at $u = u_0$ the system starts to
radiate ($c_u \neq 0$) until $u = u_f$, when the news vanish again. For 
$u>u_f$
the system is not radiating although (in general) $M_{\theta} \neq 0$ 
implying
(see for example (\ref{N})) time dependence of metric functions
(non-radiative motions \cite{Boal}).

For $u>u_f$ there is a vorticity term of order $1/r^2$
describing the effect of the tail of the wave. This in turn
provides ``observational'' evidence for the violation of the Huygens's
principle, a problem largely discussed in the literature (see for example
\cite{Boal, BoNP},
  and references therein).

\section{The electric and magnetic  parts of Weyl tensor}
The electric and magnetic parts of Weyl tensor, $E_{\alpha \beta}$ and
$H_{\alpha\beta}$, respectively, are formed from the Weyl tensor $C_{\alpha
\beta \gamma \delta}$ and its dual
$\tilde C_{\alpha \beta \gamma \delta}$ by contraction with the four
velocity vector given by (\ref{fvct}) \cite{Mac}:
\begin{equation}
E_{\alpha \beta}=C_{\alpha \gamma \beta \delta}u^{\gamma}u^{\delta}
\label{electric}
\end{equation}
\begin{equation}
H_{\alpha \beta}=\tilde C_{\alpha \gamma \beta \delta}u^{\gamma}u^{\delta}=
\frac{1}{2}\epsilon_{\alpha \gamma \epsilon \delta} C^{\epsilon
\delta}_{\quad \beta \rho} u^{\gamma}
u^{\rho},
\qquad \epsilon_{\alpha \beta \gamma \delta} \equiv \sqrt{-g}
\;\;\eta_{\alpha \beta \gamma \delta}
\label{magnetic}
\end{equation}
where
$\eta_{\alpha\beta\gamma\delta}= +1$ for $\alpha, \beta, \gamma, \delta$ in
even order, $-1$ for $\alpha, \beta, \gamma, \delta$ in
odd order and $0$ otherwise.
Also note that

$$  \sqrt{-g} = r^2 \sin \theta e^{2\beta} \approx r^2 \sin \theta
\exp{(-\frac{c^2}{2r^2})} \approx r^2 \sin \theta + O(1)$$

Since the obtained expressions are fairly long, in order to check
our results we have calculated the magnetic and electric parts in
two different ways and then compared results, excluding thereby any
possible error. On the one hand we have calculated with Maple the
components of Weyl tensor and from them, ``by hand'' its electric
and magnetic part from (\ref{electric}) and (\ref{magnetic}). We do
not include here the detailed expressions of Weyl components, but
they are available upon request. The only non--vanishing components
of Weyl tensor are $$C_{0101}, C_{0102}, C_{0112}, C_{0202},
C_{0212}, C_{0303}$$  $$C_{0313},  C_{0323},  C_{1212},  C_{1313},
C_{1323}, C_{2323}.$$ However they are not independent, since the
following relations between them exist:
\begin{eqnarray}
\frac{r^4\sin^2\theta}{e^{2\beta}}C_{1010}=e^{2\gamma}(V-r^4e^{2\gamma-2\beta})
C_{1313}-2r^2e^{2\gamma}C_{0313}, \label{14} \\
\frac{r^2\sin^2\theta}{e^{2\gamma}}C_{0112}=e^{2\beta}C_{1323}-r^2e^{2\gamma
}C_{1313},
\label{15}\\
\frac{2r^2\sin^2\theta}{e^{2\gamma}}C_{0212}=e^{2\beta}C_{2323}-re^{2\gamma}
C_{1313},
\label{16} \\
\sin^2\theta C_{1212}=-e^{4\gamma}C_{1313}.
\end{eqnarray}

On the other hand, we have calculated the magnetic and electric part using
GRTensor.

Thus, one has for the components of the magnetic Weyl tensor, up to the
order  $1/r^3$:

\begin{equation}\label{a1m}
     H^0_{0} = H^0_{1} = H^0_{2} = H^1_0 =H^1_1 = H^1_2=H^2_0 =H^2_1 =H^2_2
=H^3_0 =H^3_3 =  0
\end{equation}

$$
H^0_{3} = -\frac{1}{r}\left(2c_u \cos\theta + c_{\theta u}\sin\theta
\right)
$$
$$
   + \frac{1}{r^2}\left\{ 4 c_u(c-M) \cos\theta  +  \left[
\frac{3}{2}(N_u+M_\theta+c_u c_\theta) +  \frac{7}{2}c c_{\theta u}
   - 2M c_{\theta u}\frac{ }{ }\right]\sin\theta \right\}
$$
$$
+ \frac{1}{r^3}\left\{-\frac{N}{\sin\theta}(1+2c_u) +
\left[8Mc_u(c-M) + N_\theta (1-2c_u) +\frac{5}{2} c^2c_u-Nc_{\theta
u}- P_u -4Mc\right]\cos\theta  \right.
$$
$$
   + \left[2(N-Mc_\theta)- c_{\theta u}( 7Mc-4M^2-N_\theta-\frac{7}{4}c^2) +
3M(N_u +M_\theta)-\frac{1}{2}P_{\theta u} \right.
$$
\begin{equation}
  \left. \left. -3cM_\theta + N_{\theta\theta} + c_u(8N + 3M c_\theta +\frac{5}{2} c c_\theta )\right]\sin\theta  \right\},\label{a2m}
\end{equation}

$$
H^1_{3} = \frac{1}{r}\left[(c_{\theta}c_{uu} + c_{\theta
u})\sin\theta + 2(cc_{uu} + c_u)\cos\theta \right] + \frac{1}{r^2}
\left\{ \frac{4cc_u \cos\theta}{\sin^2\theta} + \frac{2c_u
c_{\theta} -c c_{\theta u}}{\sin\theta}\right.
$$
$$
\left. + \left[ -\frac{1}{2}c_{\theta} c_{\theta \theta u} +
(2M-3c)c_{\theta} c_{uu} - \frac{5}{2}(cc_{\theta u}+c_uc_{\theta})
-2N c_{uu} - \frac{3}{2}(N_u + M_\theta)\right]\sin\theta \right.
$$
$$
\left. + \left[ -6cc_u + 4c(M-c) c_{uu} -\frac{1}{2}c_{\theta}
c_{\theta u} - c c_{\theta \theta u}\frac{}{} \right]\cos\theta
\right\}
$$
$$
+\frac{1}{r^3} \left\{\frac{8cc_u(M-c)\cos\theta}{\sin^2\theta}+
\frac{1}{\sin\theta}\left[ c_u c_{\theta} (4M-3c) + N -2cN_u +2Nc c_{uu} +
\right.\right.
$$
$$
\left.\left. + c c_{\theta u} (7c-2M)-4c_uN-cM_\theta \right]+  \left[
\frac{1}{2}P_{\theta u} -N_{\theta \theta} -2N +4cM_\theta -4c_uN +2cN_u
\right. \right.
$$
$$
\left.\left. + c_u c_\theta(-\frac{9}{2} c-M +cc_u
+\frac{1}{2}c_{\theta\theta}) + c_\theta (2M -cM_u+P_{uu} +
N_{\theta u} +\frac{1}{2}M_{\theta\theta})   \right.\right.
$$
$$
\left.\left. + c_{\theta u}(-\frac{19}{4} c^2 + 2c_\theta^2 + 2Mc) +
c_{\theta\theta u}(N+3cc_\theta-Mc_\theta) \right.\right.
$$
$$
\left. + c_\theta c_{uu}(-6Mc+\frac{1}{2}c^2+N_\theta+4M^2)
-2Nc_{uu} (c+2M) \right] \sin\theta
$$
$$
\left.+ \left[2c(2M+N_{\theta
u}+P_{uu}-cM_{u}+\frac{1}{2}M_{\theta\theta}) +c_u(-2Mc -\frac{3}{2}
c^2 +2c^2 c_u + \frac{3}{2}c_\theta^2 +c c_{\theta\theta})
\right.\right.
$$
$$
\left.\left. + c_{\theta u}(N-Mc_\theta +8cc_\theta) + c_{uu}(c^3
+2cN_\theta+ Nc_\theta -8Mc^2+8M^2c)   \right.\right.
$$
\begin{equation}
\left.\left.+ cc_{\theta\theta u}(5c-2M) -c_\theta(\frac{1}{2}
M_\theta +N_u) +P_u -N_\theta\right]\cos\theta \right\}, \label{a3m}
\end{equation}

$$
H^2_3 = \frac{1}{r}\left(\sin\theta c_{uu}\right) +
\frac{1}{r^2}\left\{\frac{2c_u}{\sin\theta} - \frac{1}{2}c_{\theta
u}\cos\theta + \left[2c_{uu}(M-c)-c_u-\frac{1}{2} c_{\theta\theta
u}\right]\sin\theta\right\}
$$
$$
+\frac{1}{r^3}\left\{4c_u \frac{M-c}{\sin\theta} +
\left[\frac{3}{2}c_uc_\theta-N_u-\frac{1}{2} M_\theta + N c_{uu}
+(\frac{7}{2}c-M)c_{\theta u} \right]\cos\theta  + \right.
$$
$$
\left. \left[(\frac{5}{2}c-M)c_{\theta\theta u} +
(\frac{1}{2}c_{\theta\theta}-M)c_u +2c_\theta c_{\theta u}
+c_{uu}(2c^2+4M^2-4Mc+N_\theta) + \right.\right.
$$
\begin{eqnarray}
\left.\left.\frac{1}{2}M_{\theta\theta} - cM_u + N_{\theta
u}+P_{uu}+cc_u^2 \right]\sin\theta\right\},\label{a4m}
\end{eqnarray}
\begin{eqnarray}
H^3_1 = \frac{1}{r^3} \frac{2c_u\cos\theta + c_{\theta
u}\sin\theta}{\sin^2\theta},\label{a5m}
\end{eqnarray}

$$
H^3_2 = \frac{1}{r}\frac{c_{uu}}{\sin\theta} +
\frac{1}{r^2\sin\theta}\left[2c_{uu} (c+M) - c_u -\frac{1}{2}
c_{\theta\theta u} -\frac{1}{2} \cot \theta c_{\theta u} +
2\frac{c_u}{\sin^2\theta} \right]
$$
$$
+\frac{1}{r^3\sin\theta}\left\{ \frac{4Mc_u}{\sin^2\theta} +\cot
\theta \left[-N_u + Nc_{uu} - \frac{1}{2} c_uc_\theta -(\frac{1}{2}
c +M)c_{\theta u} -\frac{1}{2}M_\theta \right]\right.
$$
$$
\left. + c_{uu}(N_\theta + 4Mc +4M^2 +2c^2) +cc_u^2 +
(\frac{1}{2}c_{\theta\theta}-M)c_u + (\frac{1}{2}c-M)
c_{\theta\theta u}\right.
$$
\begin{equation}
\left.+ P_{uu} + \frac{1}{2}M_{\theta\theta} + c_\theta c_{\theta u}
- cM_u + N_{\theta u} \right\}.\label{a6m}
\end{equation}



\noindent Regarding the electric part, one gets, up to the  order $1/r^3$:

\begin{equation}\label{1e}
     E^{0}_{0} = E^{0}_{3} = E^{1}_{0} = E^{1}_{3} = E^{2}_{0}
     = E^{2}_{3} = E^{3}_{0} = E^{3}_{1} = E^{3}_{2}
     = 0
\end{equation}
\begin{equation}\label{2e}
      E^{0}_{1} = \frac{2(cc_u + M)}{r^3},
\end{equation}

$$
      E^{0}_{2} = \frac{2c_u \cos\theta + c_{\theta u} \sin\theta}{r
      \sin\theta}
$$
$$ +\frac{1}{2\sin\theta r^2}\left\{8Mc_u
      \cos\theta + \left[ c_{\theta u} (4M-3c) -3 (M_\theta +c_uc_\theta
+N_u) \right]\sin\theta \right\}
$$
$$
      +\frac{1}{4 r^3}\left\{ \left(1+2c_u\right)\frac{4N}{\sin^2\theta} +
\cot\theta
      \left[4 \left(Nc_{\theta u}+P_u-N_\theta\right) + c_u \left( 32M^2
+8N_\theta-42 c^2\right)\right]\right.
$$
$$
      \left. + 4 \left( N-3cN_u -N_{\theta\theta}  \right) -12M
\left(M_\theta + N_u \right)  + c_{\theta u}\left(4N_\theta + 16M^2-13
      c^2-12 Mc \right)\right.
$$
\begin{equation}
      \left. -   c_u\left( 30 c c_\theta +12 M c_\theta + 32 N  \right) +
2P_{\theta
      u} \frac{}{}
      \right\}, \label{3e}
\end{equation}
\begin{equation}\label{4e}
      E^{1}_{1} = -\frac{2cc_u(1+\cos^2\theta) + (c_\theta c_{\theta u}  +
2M)\sin^2\theta + 2\sin\theta\cos\theta (c_u c_\theta + c c_{\theta
u})}{r^3 \sin^2\theta},
\end{equation}

$$
      E^{1}_{2} = -\frac{2(cc_{uu} + c_u)\cos\theta + (c_\theta c_{u u}  +
c_{\theta u})\sin\theta}{r \sin\theta}
$$
$$
      + \frac{1}{2 r^2} \left\{ \frac{}{}3(N_u +M_\theta) +
      c_{uu} (2cc_\theta -4Mc_\theta +4N ) + 5c_\theta c_u
      + c_\theta c_{\theta\theta u} + c c_{\theta u} \right.
$$
$$
      \left. + \cot\theta \left[ 2c ( c_{\theta\theta u} +2 c_u - 4M c_{uu})
+ c_\theta c_{\theta
      u}\right]  + \frac{2}{\sin^2\theta}(cc_{\theta u}-2 c_\theta c_u)
      -\frac{8}{\sin^3\theta} c c_u \cos\theta \right\}
$$
$$
      - \frac{1}{r^3}\left\{8cc_u (M-c)\frac{\cos\theta}{\sin^3\theta} +
\frac{1}{\sin^2\theta}\left[(4M-7c)c_uc_\theta-4c_uN+
      2Ncc_{uu}+(c^2-2Mc)c_{\theta u}\right.\right.
$$
$$ \left.+ N-c(M_\theta+2N_u) \frac{}{}\right]+\cot\theta\left[
c_u(-\frac{1}{2}c_\theta^2-2Mc+cc_{\theta\theta}+2c^2c_u-\frac{15}{2}c^2)+
\right.
$$
$$
\left. c_\theta\left(c_{uu}N +(3c-M)c_{\theta u} -N_u
-\frac{1}{2}M_\theta \right) + cc_{uu}(8M^2-3c^2+2N_{\theta}) +
cc_{\theta\theta u}(3c-2M) \right.
$$
$$
\left.+c_{\theta u} N + cM_{\theta\theta} + P_u - N_\theta +
2c(P_{uu}+N_{\theta u}-M-cM_u)\frac{}{} \right]
$$
$$
+ c_{uu}\left[c_\theta(-\frac{7}{2}c^2
+N_\theta-2Mc+4M^2)-6Nc-4MN\right] + c_u\left[
c_\theta(\frac{1}{2}c_{\theta\theta}-\frac{9}{2}c-M+cc_u)-4N\right]
$$
$$
+c_{\theta u} (c_\theta^2+2Mc -\frac{15}{4}c^2)+c_{\theta\theta
u}(N-Mc_\theta +
2cc_\theta)+c_\theta(\frac{1}{2}M_{\theta\theta}-M+N_{\theta u} +
P_{uu} -M_u c)
$$
\begin{equation}\label{5e}
   \left. -N_{\theta\theta} + \frac{1}{2}P_{\theta u} - cN_u + N
   +cM_\theta\right\},
\end{equation}

\begin{equation}\label{6e}
      E^{2}_{1} = -\frac{2 c_u \cos\theta +  c_{\theta u} \sin\theta}{r^3
\sin\theta},
\end{equation}

$$
      E^{2}_{2} = -\frac{  c_{uu} }{r} +
      \frac{1}{2r^2}\left(c_{\theta\theta u}-4Mc_{uu} + 2c_u +
      \cot\theta c_{\theta u}- \frac{4c_u}{\sin^2\theta}\right)
$$
$$
+\frac{1}{r^3}\left\{M_uc-\frac{1}{2}M_{\theta\theta}+M-N_{\theta
u}-P_{uu} \right.
$$
$$
  + \cot\theta \left[Mc_{\theta u} + \frac{1}{2}M_\theta
+N_u-\frac{1}{2}cc_{\theta u} + \frac{1}{2}c_\theta c_u-
Nc_{uu}\right]
$$
\begin{equation}\label{7e}
\left. + c_u\left[\frac{4(c-M)}{\sin^2\theta}
+M-c-cc_u-\frac{1}{2}c_{\theta\theta}
\right]-c_{uu}(4M^2+N_\theta)-c_\theta c_{\theta u} +
c_{\theta\theta u}(M-\frac{3}{2}c)\right\},
\end{equation}

$$
      E^{3}_{3} = \frac{ c_{uu} }{r}-\frac{1}{2r^2}\left(c_{\theta\theta
u}-4Mc_{uu} + 2c_u +
      \cot\theta c_{\theta u}- \frac{4c_u}{\sin^2\theta}\right)
$$
$$
+\frac{1}{r^3}\left\{M+N_{\theta u} + P_{uu} +
\frac{1}{2}M_{\theta\theta}-cM_u  \right.
$$
$$
+ \cot\theta \left[ \frac{5}{2} cc_{\theta u} -
\frac{1}{2}M_\theta-N_u +Nc_{uu}-Mc_{\theta u} + \frac{3}{2}c_u
c_\theta \right]
$$
\begin{equation}\label{8e}
\left. + c_u\left[\frac{4M}{\sin^2\theta}+cc_u -M
+\frac{1}{2}c_{\theta\theta}-c\right] + c_{uu} (4M^2+N_\theta) +
2c_\theta c_{\theta u} + (\frac{3}{2} c-M) c_{\theta\theta
u}\right\},
\end{equation}
where
\begin{equation}
P\equiv C-\frac{c^3}{6}.
\label{P}
\end{equation}

  We shall now provide expressions for the two algebraic
invariants associated to the electric and magnetic parts of the Weyl
tensor, namely
\begin{equation}\label{invariants}
     Q= H^\alpha_{\; \beta} E^\beta_{\;\alpha}, \qquad  L = E^\alpha_{\; 
\beta}
     E^\beta_{\;\alpha}- H^\alpha_{\; \beta} H^\beta_{\;\alpha}.
\end{equation}

As it turns out, the invariant $Q$ vanishes identically; i.e.:

\begin{equation}\label{M}
     Q = 0
\end{equation}

whereas the first non-vanishing order in $L$  is $1/r^6$ and one
then has:

\begin{equation}\label{L}
     L = \frac{2}{r^6}\left[3(cc_u +M)^2 + (c^3+6P) c_{uu} +6N (
     c_{\theta u} + 2 c_u \cot\theta)
    \right] + O(1/r^7)
\end{equation}

\section{Discussion}

We are now ready to try to answer to the main questions which motivated
this work, in the context of the Bondi metric, namely:
\begin{itemize}
\item What consequences do emerge from the vanishing of the magnetic part
of the Weyl tensor?
\item What consequences do emerge from the vanishing of the electric part
of the Weyl tensor?
\item How do different types of fields (radiative, non radiative but time
dependent, and static ) enter into the electric and magnetic part of the
Weyl tensor, and into the
corresponding invariants?
\item Why does gravitational radiation produce vorticity?
\end{itemize}

Let us start with the first question. If we put $H^{\alpha}_{\beta}=0$ then
it follows from the coefficient of $1/r$  in (\ref{a2m}) that
\begin{equation}
c_{\theta u} \sin \theta +2c_{u}\cos \theta=0
\label{1}
\end{equation}
which according to (\ref{cu}) and (\ref{F}) implies
\begin{equation}
c_{u}=0.
\label{n2}
\end{equation}
Thus the field is non--radiative.
Next, the vanishing of the coefficient $1/r^2$ in (\ref{a2m}) implies in
turn that
\begin{equation}
M_{\theta}=N_{u}=0,
\label{n3}
\end{equation}
  where we have used
(\ref{N}).

  Finally from the vanishing of the coefficient of the $1/r^3$ in
(\ref{a2m}) it follows that
\begin{equation}
N_{\theta \theta} \sin^2 \theta +N_{\theta}\sin \theta \cos \theta-N \cos
2\theta-(2cM\sin^2 \theta)_{\theta}=0
\label{3}
\end{equation}
whose general solution is
\begin{equation}
N=\left(\int \frac{2cM}{\sin \theta}\,d\theta +\sigma \right)\sin\theta
\label{4}
\end{equation}
where $\sigma $ is a constant.

Feeding back (\ref{4}) into (\ref{C}) and using (\ref{n2}), it follows that
\begin{equation}
C_{u}=0.
\label{5}
\end{equation}
It can be easily checked that no further information can be obtained from
(\ref{a3m}--\ref{a6m}).
Therefore up to order $1/r^3$ in $\gamma$, the metric is static, and the
mass, the ``dipole'' ($N$) and the ``quadrupole'' ($C$) moments correspond
to a satic situation. However, the
time dependence might enter through coefficients of higher order in
$\gamma$, giving rise to what Bondi calls ``non--natural non--radiative
moving system'' (nnnrms). In this later case,
the system keeps the three first moments independent of time, but allows
for time dependence of higher moments. As unlikely as this situation may be
from the physical point of view, we
were not able  not rule it out mathematically. On the other hand it is
known that static  space--times are purely electric. Accordingly, we
conclude that, exluding
nnnrms, the necessary and sufficient condition for a  Bondi metric to be
purely electric is to be static.

The second question has a simple answer. Indeed assuming
$E^{\alpha}_{\beta}=0$ and using regularity conditions, we find from the
coefficient of order $1/r$ in (\ref{3e})
\begin{equation}
c_{u}=0,
\label{1elec}
\end{equation}
then, it follows at once from (\ref{2e}) that $M=0$. If we exclude the
possibility of negative masses, then the spacetime must be Minkowski,
giving further support to the
conjecture that there are no purely magnetic vaccum space--times 
\cite{Bonnor}.

The third question has also a simple answer. The electric part contains all
kinds of contributions, radiative, non--radiative but time dependent
(nrtd), and  static. At order $1/r$ only
radiative contributions appear, whereas nrtd terms appear  at order
$1/r^2$, and higher and contributions from the static field enter in the
$1/r^3$ order, and higher. However the
magnetic part does not contain contribution from the static field. Only radiative (at order $1/r$  and higher) and nrtd terms (at order $1/r^2$  and higher) appear.

On the other hand,
the only non--vanishing invariant ($L$) has a leading term of order $1/r^6$ implying that purely radiative  and nrtd terms in the Weyl tensor do not contribute to $L$. This is in
agreement with the fact that for purely radiative spacetimes, both invariants $Q$ and $L$ vanish \cite{Bonnora}

Finally, let us consider the last question.  With this purpose in mind, it
is worth recalling a result obtained by Bonnor \cite{bonleter} concerning
the dragging of inertial frames by a
charged magnetic dipole. To explain the appearance of vorticity in such
space--times, Bonnor notices that the corresponding electromagnetic
Poynting vector has a non--vanishing
component, describing a flow of electromagnetic energy round in circles
where frame--dragging occurs \cite{Feymann}. He then suggests that such a
flow of energy affects inertial frames
by producing vorticity of congruences of particles, relative to the compass
of inertia.

One could speculate about a similar mechanism in our case, i.e. a flow of
gravitational radiation in the $\phi$ direction. However for testing such a
conjecture we should have
available a unique expression for a ``gravitational'' Poynting vector,
which is still an open question in general relativity.

Thus for example, the  super--Poynting vector  based on the Bel--Robisnson tensor, as defined in \cite{basset}, is 
\begin{equation}
P_{\alpha}=\epsilon_{\alpha \beta \gamma \delta}E^{\beta}_{\rho}H^{\gamma \rho}u^{\delta}
\label{p1}
\end{equation}
giving in our case $P^{\phi}=0$. However, besides the ambiguity problem in the definition of energy, this negative result may be caused by the reflection symmetry of the Bondi
metric, which intuitively seems to be incompatible with the presence of a circular flow of energy in the $\phi$ direction. In order to clarify this situation, $P^{\phi}$ should be
calculated for the general radiative metric without reflection symmetry \cite{Sachs}, but this of course is out of the scope of the present work.


\begin{thebibliography}{88}
\bibitem{Bel} L Bel {\it Cah. de Phys.} {\bf 16} 59 (1962);  {\it Gen. Rel.
Grav.} {\bf 32} 2047 (2000)

\bibitem{glass} E N Glass {\it J. Math. Phys.} {\bf 16} 2361 (1975)

\bibitem{Barnes} A Barnes and R Rowlingson {\it Class. Quantum Grav.} {\bf
6} 949 (1989)

\bibitem{Matarrese} S Matarrese, O Pantano and D Saez {\it Phys. Rev. D}
{\bf 47} 1311 (1993)

\bibitem{Bruni} M Bruni, S Matarrese and O Pantano {\it Astrophys. J} {\bf
445} 958 (1995)

\bibitem{Wade} C McIntosh, R Arianrhod, S Wade and C Hoenselaers {\it
Class. Quantum Grav.} {\bf 11} 1555 (1994)


\bibitem{Hadow} B M Haddow {\it J. Math. Phys.}  {\bf 18} 1378 (1995)

\bibitem{Bonnora} W B Bonnor {\it Class. Quantum. Grav.}  {\bf 12}  499 
(1995)

\bibitem{Bonnor} W B  Bonnor  {\it Class. Quantum. Grav.}  {\bf 12} 1483 
(1995)

\bibitem{Dunsby} P K S Dunsby, B A Basset and G F R Ellis {\it Class.
Quantum. Grav.}  {\bf 14} 1215 (1997)

\bibitem{Maartens} R Maartens, G F R Ellis and T Siklos {\it Class.
Quantum. Grav.}  {\bf 14} 1927 (1997)

\bibitem{Hogan} P Hogan and G F R Ellis  {\it Class. Quantum. Grav.}  {\bf
14} A171 (1997)

\bibitem{van} H van Elst, C Uggla, W M  Lesame, G F R Ellis and R Maartens 
{\it Class. Quantum Grav} 1, 1151 (1997).

\bibitem{vanII} H van Elst   and G F R Ellis {\it  Class. Quantum
Grav.}  15 3545 (1998).

\bibitem{basset} R Maartens and B Basset {\it Class. Quantum Grav.}
{\bf 15} 705 (1998).

\bibitem{Lesame}  R Maartens, W M Lesame and  G F R Ellis {\it Class.
Quantum. Grav.}  {\bf 15} 1005 (1998)

\bibitem{Loz} C Lozanowski and M Aarons {\it Class. Quantum. Grav.}  {\bf
16} 4075 (1999)

\bibitem{Bergh} N Van den Bergh {\it Class. Quantum. Grav.}  {\bf 20} L1
(2003); {\it Class. Quantum. Grav.}  {\bf 20} L165 (2003) 

\bibitem{Ferrando} J Ferrando and J Saez {\it Class. Quantum. Grav.}  {\bf
20} 2835 (2003)

\bibitem{HHP} L  Herrera and J L Hernandez-Pastora, {\it Class. Quantum.
Grav.} {\bf 17} 3617 (2000)

\bibitem{Valiente} J Valiente  {\it Class. Quantum.Grav.} {\bf 18} 4311 
(2001)

\bibitem{Felice} F Sorge, D Bini, F de Felice {\it Class. Quantum Grav.}
{\bf 18} 2945 (2001)

\bibitem{bonleter} W B Bonnor {\it Phys. Lett. A} {\bf 158} 23 (1991)

\bibitem{Boal} H Bondi, M G J  van der Burg and A W K  Metzner
{\it Proc. R. Soc.} {\bf A269} 21 (1962)

\bibitem{JaNe} A J Janis and E T  Newman  {\it J.Math.Phys.} {\bf 6} 902 
(1965)

\bibitem{BoNP} W B  Bonnor in {\it Proceedings of the Meeting on General
Relativity}
Florence 1965, ed. G. Barbera (Florence, 1965), Vol.II;
E T  Newman and R  Penrose, {\it Proc. R. Soc.} {\bf A305} 175 (1968)


\bibitem{Bo} W B  Bonnor in {\it Ondes et Radiations Gravitationelles}
(Editions du
Centre National de la Recherche Scientifique, Paris) (1974)


\bibitem{Mac} H Stephani, D Kramer, M MacCallum,C Honselaers and
E  Herlt, {\it Exact Solutions to Einstein's Field Equations. Second 
Edition},
(Cambridge University Press, Cambridge), (2003)

\bibitem{Feymann} R P Feynman, R B Leighton and M Sand {\it Lectures on
Physics} {\bf II} (Addison--Wesley, Reading, 1964) pp. 27, 28

\bibitem{Sachs} R Sachs 
{\it Proc. R. Soc.} {\bf A270} 103 (1962)


\end{thebibliography}
\end{document}